\documentclass[submission, Phys]{SciPost}

\usepackage{graphicx}
\usepackage{dcolumn}
\usepackage{bm}

\begin{document}

\sloppy

\begin{center}{\Large \textbf{
Fingerprints of hot-phonon physics in time-resolved correlated quantum lattice dynamics
}}\end{center}

\begin{center}
E. Cappelluti\textsuperscript{1*},
D. Novko\textsuperscript{2,3}
\end{center}

\begin{center}
{\bf 1} Istituto di Struttura della Materia, CNR (ISM-CNR), 34149 Trieste, Italy
\\
{\bf 2} Institute of Physics, 10000 Zagreb, Croatia
\\
{\bf 3} Donostia International Physics Center (DIPC), 20018 Donostia-San Sebasti\'an, Spain
\\
* emmanuele.cappelluti@ism.cnr.it
\end{center}

\begin{center}
\today
\end{center}


\section*{Abstract}
{\bf
The time dynamics of the
energy flow from electronic to lattice degrees of freedom
in pump-probe setups could be strongly affected by the presence
of a hot-phonon bottleneck, which can sustain longer
coherence of the optically excited electronic states.
Recently, hot-phonon physics has been experimentally
observed and theoretically described in MgB$_2$,
the electron-phonon based superconductor with $T_{\rm c}\approx 39$ K.
By employing a combined ab-initio and a semiclassical approach and by taking MgB$_2$ as an example, here
we propose a novel path for revealing the presence and characterizing the properties of hot phonons through a direct analysis
of the information encoded in the lattice inter-atomic correlations.
Such method exploits the underlying symmetry of the
$E_{2g}$ hot modes characterized by a out-of-phase
in-plane motion of the two boron atoms.
Since hot phonons occur typically at high-symmetry
points of the Brillouin zone, with specific symmetries
of the lattice displacements,
the present analysis is quite general
and it could aid in revealing the hot-phonon physics in other promising materials, such as graphene, boron nitride, black phosphorus, or even underdoped cuprates.
}

\vspace{10pt}
\noindent\rule{\textwidth}{1pt}
\tableofcontents\thispagestyle{fancy}
\noindent\rule{\textwidth}{1pt}
\vspace{10pt}

\section{Introduction}
\label{s:intro}

Ultrafast time-resolved measurements have proven in the past few decades
to be a powerful tool for investigating fundamental mechanisms of a variety of physical phenomena
and for revealing processes governing different states of the matter, where several degrees of freedom
(e.g., electrons, lattice and magnetic modes) play a mutual role\,\cite{bib:perfetti07,ishioka08,yan09,gadermaier10,mischa11,cortes11,rudolf12,arnaud13,bib:johannsen13,jensen14,chase16,waldecker16,cotret19}.
In a typical pump-probe setup, energy is initially injected into the electronic degrees of freedom
by means of particle-hole excitations, giving rise to a non-thermal electronic distribution.
The time evolution of such non-thermal excitations is governed by different scattering mechanisms
that lead to an energy transfer towards other degrees of freedom, in particular in the lattice mode sector,
until, on a long time scale, a new final steady state is reached where all the different degrees of freedom
are thermalized and at equilibrium.

The possibility of sustaining non-thermal states for sufficiently long time opens striking opportunities
in the field of quantum information as long as the scattering sources affecting such states
can be kept under control. The electron-phonon (el-ph) coupling is of major importance in this context
since it represents one of the primary channels responsible for the internal thermalization of the electronic
degrees of freedom, and because it leads eventually, when it is not constrained, to a system heating detrimental for any coherent process\,\cite{bib:sentef13}.

However, the thermalization process of some materials can be hindered by the so-called phonon bottleneck\cite{price85,kocevar85,potz87,joshi89,langot96,butscher07,ishioka08,yan09,kang10,berciaud10,lui10,huang11,scheuch11,wu2012,hannah13,golla17,yang16,koi17,hamham18,chan21,sekiguchi21,tong21,seiler21} that quenches
the energy transfer between electron and lattice degrees of freedom.
Such a physical phenomenon is usually encountered in semiconductors and semimetals where
the pump-driven particle-hole excitations are restricted to few single points (valleys)
in the Brillouin zone. In such conditions only a few phonons with the right symmetry
and with selected momentum ${\bf q}$ connecting two valleys can be excited efficiently.
Energy from the electron sub-system is thus prevalently transferred to these modes
that get ``hot'', meaning they acquire a phonon population $b_{\bf q}$ much larger
than other modes, giving rise to a non-thermal phonon distribution.

Within this context, the possibility of a phonon bottleneck, and hence of hot phonons, 
has been so far associated only to semiconductors and semimetals, whereas conventional metals,
with a large Fermi surface allowing scattering with many ${\bf q}$-phonons, were thought to be incompatible
with such a scenario.
At odds with this belief,  a novel path for inducing hot phonons has been recently proposed
for unconventional metals characterized by a strong anisotropy of the el-ph coupling \cite{baldini,ncdc}.
In the paradigmatic case of MgB$_2$, the el-ph coupling is concentrated in a few in-plane modes at the Brillouin zone center
possessing the $E_{2g}$ symmetry\cite{bib:liu01,an,yildirim,bib:choi02b,bib:choi02a,kong,golubov,gonnelli},
with a strong resemblance to the relevant modes in single- and multi-layer graphene.
As a consequence of such remarkable anisotropy, the energy initially pumped into the electron sector is efficiently and rapidly ($\sim 50$ fs) transferred only to such few lattice degrees of freedom that get a much larger population compared to the remaining lattice modes,
whereas a final thermalization among the whole phonon modes and among electron and lattice degrees of freedom occurs on a much larger time scale ($\sim 0.4$ ps)\,\cite{ncdc}.
On the experimental ground,
an indirect signature of such hot-phonon scenario has been observed in the onset of unconventional features
in the time-resolved optical spectra \cite{baldini}.
More direct  and precise fingerprints have been proposed at the theoretical level in the analysis
of time-resolved Raman spectroscopy, both in spectral features as well as in integrated spectral weights \cite{ncdc}.
An accurate measurement of time-resolved Raman spectra in MgB$_2$ needs to face
with the limitations of the uncertainty principle constraining time and energy resolution \cite{versteeg}.
In addition, it should be remarked that in both cases the presence of a non-thermal population of the in-plane $E_{2g}$ modes
is not directly probed via the properties of the lattice degrees of freedom, but indirectly
through the el-ph driven many-body renormalization of the electronic response, i.e.
in the optical response \cite{baldini} or in the phonon self-energy \cite{ncdc,novko18}, which is strictly related to the electronic screening.

In the present work we suggest that a suitable evidence of the presence of hot phonons
in MgB$_2$ can be attained in a more straightforward way by the analysis of the time-resolved
lattice dynamics, as obtained for instance by means of ultrafast electron diffuse
scattering \cite{carbone08,carbone10,chatelain14,chase16,harb16,waldecker16,waldecker17,stern18,
konstantinova18,cotret19,zhan20,seiler21,otto21,Zacharias_joint_PRB,Zacharias_joint_PRL}.
In particular, we show that the inter-atomic pair-distribution function,
which encodes information about the correlated atomic motion, is highly sensitive
to the presence of a hot phonon and it provides a useful tool for detecting them.
As related to the analysis of time-independent quantities,
our proposal
is not affected by the limitations of uncertainty principle.
Note that the present analysis, here applied to the paradigmatic case of MgB$_2$,
does not rely on specific properties of this material but
it can be employed as well in other compounds, such as single-layer and multi-layer graphene\,\cite{novko2019,caruso20}, transition metal dichalcogenides\,\cite{caruso21}, black phosphorous\,\cite{seiler21,zhan20}, hole-doped diamond\,\cite{boeri04,giustino07}
and underdoped cuprates \cite{mansart13,johnson17}, where el-ph coupling plays a major role.

The paper is structured as follows: Sec. \ref{s-model} introduces the current
modelling for hot phonons in MgB$_2$ as previously discussed in Ref. \cite{ncdc}.
Such basic notions will be employed in Sec. \ref{s-latticedynam} to investigate
the effects of hot phonons on the correlated lattice dynamics in MgB$_2$.
A wider perspective of the present analysis in regards to other materials
and possible experimental approaches is finally provided in Sec. \ref{s-wider}.

\section{Theoretical modelling}
\label{s-model}

\begin{figure}[t]
\begin{center}
\includegraphics[width=0.88\textwidth]{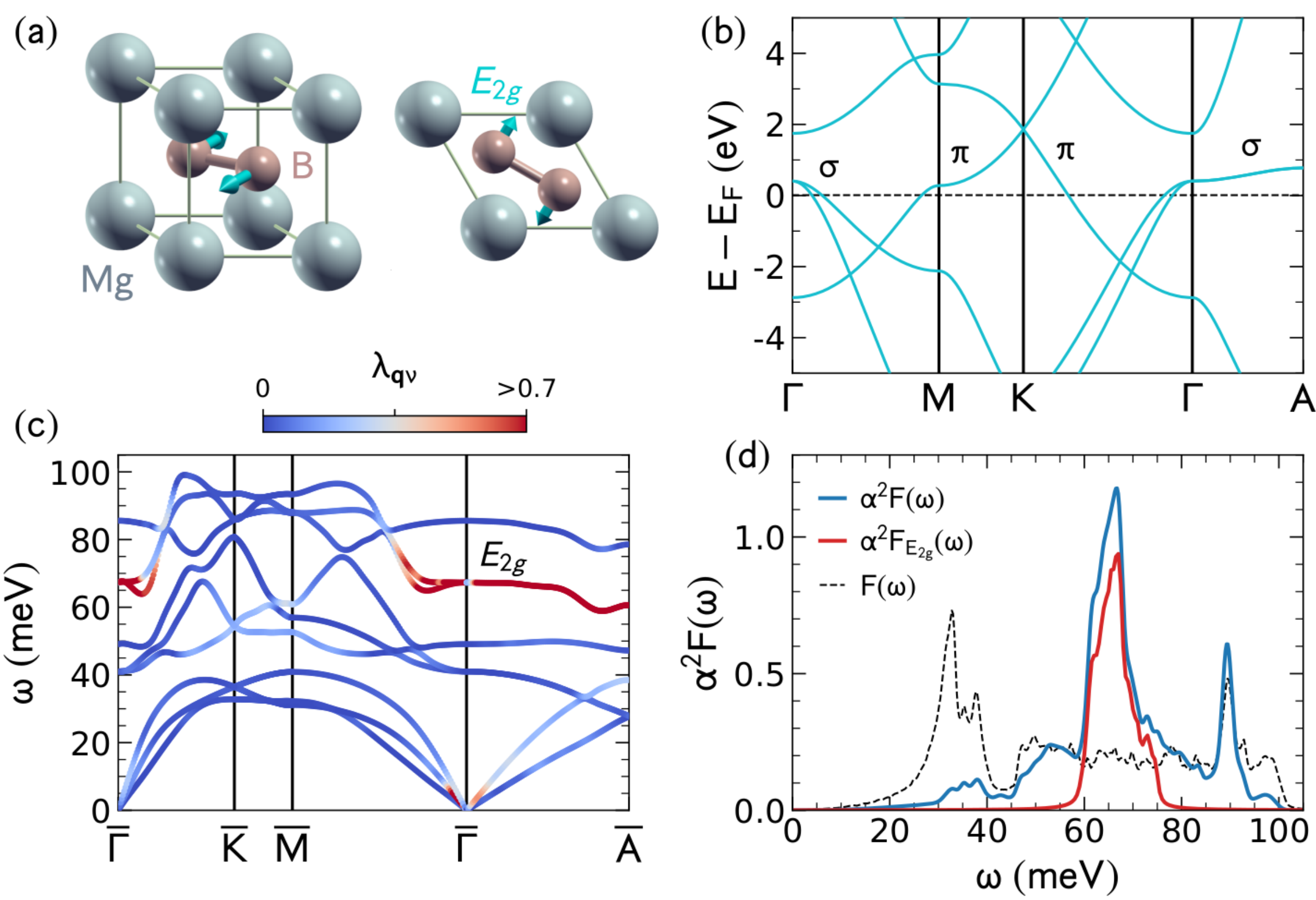}
\end{center}
\caption{
(a) Crystal structure of MgB$_2$. Also displayed are the lattice displacements
of the in-plane $E_{2g}$ mode at ${\bf q}=0$.
(b) Electronic band dispersion of MgB$_2$. Bands are labelled as $\sigma$-
and $\pi$-bands according to their symmetry.
(c) Phonon dispersions of MgB$_2$.
The color code (color bar atop) represents
the strength of the el-ph coupling  $\lambda_{\mathbf{q}\nu}$.
(d) Corresponding
phonon density of states $F(\omega)$ (dashed line)
and the total Eliashberg function $\alpha^2F(\omega)$ (blue solid line).
The solid red line shows the contribution to the Eliashberg function
associated with the hot $E_{2g}$ modes.
Readaptation from Ref. \cite{ncdc}.
}
\label{f-mgb2}
\end{figure}

The crystal structure of MgB$_2$ is rather simple,
with hexagonal graphene-like planes of B atoms spaced vertically by Mg atoms
located in the center of the hexagons
(see Fig.\,\ref{f-mgb2}a)\,\cite{bib:kortus01}.
In order to reproduce an {\em ab-initio} ground state properties of this material,
we employ here the
{\sc quantum espresso} package\,\cite{bib:qe}.
Norm-conserving pseudopotentials were employed with the 
Perdew-Burke-Ernzerhof
exchange-correlation functional \cite{bib:pbe}.
The in-plane lattice parameter and the distance between two
boron planes were set to 
3.083 \AA\, and 3.521 \AA,
respectively.
A $24 \times 24\times 24$ Monkhorst-Pack grid in
momentum space and a plane-wave cutoff energy of 60\,Ry were
used for ground-state calculations.
The phonon dispersion $\omega_{{\bf q},\nu}$ was calculated on a
$12\times12\times12$ grid using 
density functional perturbation theory\,\cite{bib:baroni01}.
The el-ph coupling strength for each momentum ${\bf q}$ and each phonon branch $\nu$
was computed
by using the {\sc epw} code \,\cite{bib:epw}
according the standard expression:
\begin{eqnarray}
\lambda_{{\bf q},\nu}
&=&
\frac{2}{N(0)N}
\sum_{{\bf k},n,m}
\frac{|g_{{\bf k}n,{\bf k+q}m}^{\nu} |^2}{\omega_{{\bf q},\nu}}
\delta(\epsilon_{{\bf k}n}-\mu)\delta(\epsilon_{{\bf k+q}m}-\mu)
,
\end{eqnarray}
where $\mu$ is the electron chemical potential, $\epsilon_{{\bf k}n}$ are the electron energies,
$\omega_{{\bf q},\nu}$ are the phonon frequencies,
$g_{{\bf k}n,{\bf k+q}m}^{\nu}$ are the el-ph matrix elements,
$N(\mu)=\sum_{{\bf k},n}\delta(\epsilon_{{\bf k}n}-\mu)$
is the electronic density of states per spin at the Fermi level,
and $N$ is
the total number of sampling points in the Brillouin zone. The delta functions $\delta(x)$ are modeled by Gaussian functions with finite broadenings.
Electron energies, phonon frequencies, and
electron-phonon coupling matrix elements were interpolated using maximally-localized Wannier functions\,\cite{bib:wannier}. 
The Eliashberg function, which shows a strength of the el-ph coupling for a particular phonon energy, was obtained on a
$40\times40\times40$ grid of electron and phonon momenta.

The well-known band structure is shown in Fig.\,\ref{f-mgb2}b.
Due to the symmetry of the systems, the in-plane $\sigma$-bands
retain a strong two-di\-men\-sion\-al character with a very weak amount of Mg,
very similar to the $\sigma$ bands of graphene, whereas the $p_z$ B-orbitals
strongly hybridize with Mg giving rise to $\pi$-bands with a strong
three-dimensional character.
As a consequence of such strong anisotropy in the electronic degrees of freedom,
the el-ph coupling results to be highly anisotropic,
as depicted in Fig. \ref{f-mgb2}c.
The el-ph coupling among states in the $\pi$ bands and among
$\sigma$ and $\pi$ states, involving $\pi$ bands with
a good metallic character, is quite weak.
On the other hand, the $\sigma$ bands appear to be only slightly doped,
resulting in a poor screening of the in-plane B-B lattice modes,
and in a corresponding large el-ph coupling between states in the $\sigma$ bands.
A further consequence of the small hole-doping of the two-di\-men\-sion\-al $\sigma$ bands
is the intrinsic restriction, obeying to the Lindhard model,
of a sizable el-ph coupling in a small momentum region $|{\bf q}_{||}| \le 2k_{\rm F}$,\cite{kong,bohnen,renker,shukla03,dastuto07}
where ${\bf q}_{||}=(q_x,q_y)$ and  $k_{\rm F}$ is the in-plane Fermi momentum of the $\sigma$ bands
which is only weakly dependent on $k_z$.

Within the above framework, 
a predominant role is played by the in-plane $E_{2g}$ phonon modes at small ${\bf q}_{||}$,
characterized by a strong el-ph coupling and
which corresponds to out-of-phase displacements of the two boron atoms within the unit cell,
as depicted in Fig.\,\ref{f-mgb2}a.
The overall strongly-coupled modes are thus limited, in a schematic way,
to only two phonon branches within the energy window $\omega_{{\bf q},\nu}\in[60:75]$\,meV,
with roughly $|{\bf q}_{||}| \le 2k_{\rm F} $ and weak dispersion along $q_z$,
amounting roughly to 5 \% of the total phonon modes, and corresponding thus
to a very small specific heat capacity.
The crucial role of the few $E_{2g}$ modes in the total coupling can be clearly pointed out
in the Eliashberg function (Fig.\,\ref{f-mgb2}d) which shows a remarkable peak
in the range $\omega_{{\bf q},\nu}\in[60:75]$\,meV, in spite of the absence of any
particular feature in the corresponding phonon density of states.

Sophisticated and powerful tools for evaluating the non-thermal time-evolution
of the coupled electron-lattice system have been
developed in the recent years \cite{tong21,seiler21,caruso21,Zacharias_joint_PRL,Zacharias_joint_PRB}.
Such techniques appear particularly useful when
in investigating the time range when effective electron and/or lattice
temperatures cannot be properly defined.
The hot-phonon physics we discuss here involves however
macroscopical quantities integrated over momenta,
where such level of accuracy is not needed and
effective-temperature models can provide as well a reliable description.
Within this context, 
following Ref.\,\cite{ncdc},
the predominance of the hot-phonon modes in
the total coupling can thus be captured
by splitting the total Eliashberg function in a {\em hot} and {\em cold} component,
$\alpha^2F(\omega)=\alpha^2F_{\rm hot}(\omega)
+\alpha^2F_{\rm cold}(\omega)$,
where $\alpha^2F_{\rm hot}(\omega)$
contains the contribution of the hot $E_{2g}$ modes close to the $\mathrm{\overline{\Gamma}-\overline{A}}$ path
in the relevant energy range $\omega \in [60:75]$\,meV (the phonon modes colored in red in Fig.\,1c),
while $\alpha^2F_{\rm cold}(\omega)$ takes into account
the other weakly coupled cold modes.
A similar splitting can be performed for the phonon density of states $F(\omega)=F_{\rm hot}(\omega)
+F_{\rm cold}(\omega)$.

Equipped with such first-principles input, we can describe the time-dynamics of the energy transfer
between the different degrees of freedom in terms
of three temperatures\,\cite{bib:allen87,bib:perfetti07,lui2010,bib:dalconte12,bib:johannsen13,ncdc,novko2019,caruso20,novko2021}, i.e.,
an electron one $T_{\rm e}$,
a hot-phonon temperature $T_{\rm hot}$, which governs the population of the $E_{2g}$-like strongly-coupled hot-phonon modes,
and the cold-lattice temperature $T_{\rm cold}$ that describes the temperature of the weakly coupled cold modes.
The time evolution of the these characteristic temperatures upon a pump pulse can be described by the set
of coupled equations
\begin{eqnarray}
C_{\rm e}
\frac{\partial T_{\rm e}}{\partial t}
&=&
S(z,t)
+
\nabla_z(\kappa\nabla_z T_{\rm e})
-
G_{\rm hot}(T_{\rm e}-T_{\rm hot})
- G_{\rm cold}(T_{\rm e}-T_{\rm cold}),
\label{eq:el}
\\
C_{\rm hot}\frac{\partial T_{\rm hot}}{\partial t}
&=&
G_{\rm hot}(T_{\rm e}-T_{\rm hot})
-
C_{\rm hot}\frac{T_{\rm hot}-T_{\rm cold}}{\tau_0},
\label{eq:e2g}
\\
C_{\rm cold}\frac{\partial T_{\rm cold}}{\partial t}
&=&
G_{\rm cold}(T_{\rm e}-T_{\rm cold})
+C_{\rm hot}\frac{T_{\rm hot}-T_{\rm cold}}{\tau_0}.
\label{eq:ph}
\end{eqnarray}
Here $C_{\rm e}$, $C_{\rm hot}$, and
$C_{\rm cold}$
are the specific heat capacities for the electron,
hot-phonon, and cold-phonon states, respectively,
and $G_{\rm hot}$/$G_{\rm cold}$
are the electron-phonon relaxation rates between electronic states
and hot/cold phonons modes.
They can be computed as
\begin{eqnarray}
C_{\rm e}
&=&
\int_{-\infty}^{\infty}
d\varepsilon N(\varepsilon)\varepsilon
\frac{\partial f(\varepsilon-\mu;T_{\rm e})}{\partial T_{\rm e}},
\label{eq:c_e}
\\
C_{\rm hot}
&=&
\int_{0}^{\infty}
d\omega F_{E_{\rm hot}}(\omega)\omega
\frac{\partial b(\omega;T_{E_{2g}})}{\partial T_{E_{2g}}},
\label{eq:c_e2g}
\\
C_{\rm cold}
&=&
\int_{0}^{\infty}
d\omega F_{\rm cold}(\omega)\omega
\frac{\partial b(\omega;T_{\rm cold})}{\partial T_{\rm cold}},
\label{eq:c_ph}
\\
G_{\rm hot}
&=&
\frac{2\pi k_B}{\hbar}
N (\mu)
\int d\Omega \Omega \alpha^2F_{E_{2g}}(\Omega),
\label{eq:ge2g}
\\
G_{\rm cold}
&=&
\frac{2\pi k_B}{\hbar}
N (\mu)
\int d\Omega \Omega \alpha^2F_{\rm cold}(\Omega).
\label{eq:gph}
\end{eqnarray}
Here $N(\varepsilon)$ is the electronic density of states,
$\mu$ the electronic chemical potential,
and $f(x;T)=1/[\exp(x/T)+1]$, $b(x;T)=1/[\exp(x/T)-1]$ are
the Fermi-Dirac and
the Bose-Einstein distribution functions, respectively.
The values obtained from the first-principles
calculations are $C_{\mathrm{e}}=90\,\mathrm{J/m^3K^2}\times T_{\mathrm{e}}$,
$C_{\rm hot}=0.13\,\mathrm{J/m^3K}$, and $C_{\mathrm{cold}}=4.1\,\mathrm{J/m^3K}$,
$G_{\rm hot}=2.8\times 10^{18}\,\mathrm{W/m^3K}$ and 
$G_{\rm cold}=3.6\times 10^{18}\,\mathrm{W/m^3K}$.
Furthermore, modelling a pump-probe setup, the term
$S(z,t)=I(t)e^{-z/\delta}/\delta$ takes into account
the energy absoption from the laser pulse
into the electronic degrees of freedom,
where $I(t)$ is the intensity of the absorbed fraction
of the laser pulse (with a Gaussian profile)
and $\delta$ is the penetration depth.
Finally, the relaxation time $\tau_0$ takes into account
the scattering between hot and cold modes driven by
the anharmonic phonon-phonon coupling, for which we take an estimate of 400\,fs\,\cite{bib:shukla03}.

\begin{figure}[!t]
\begin{center}
\includegraphics[width=0.8\textwidth]{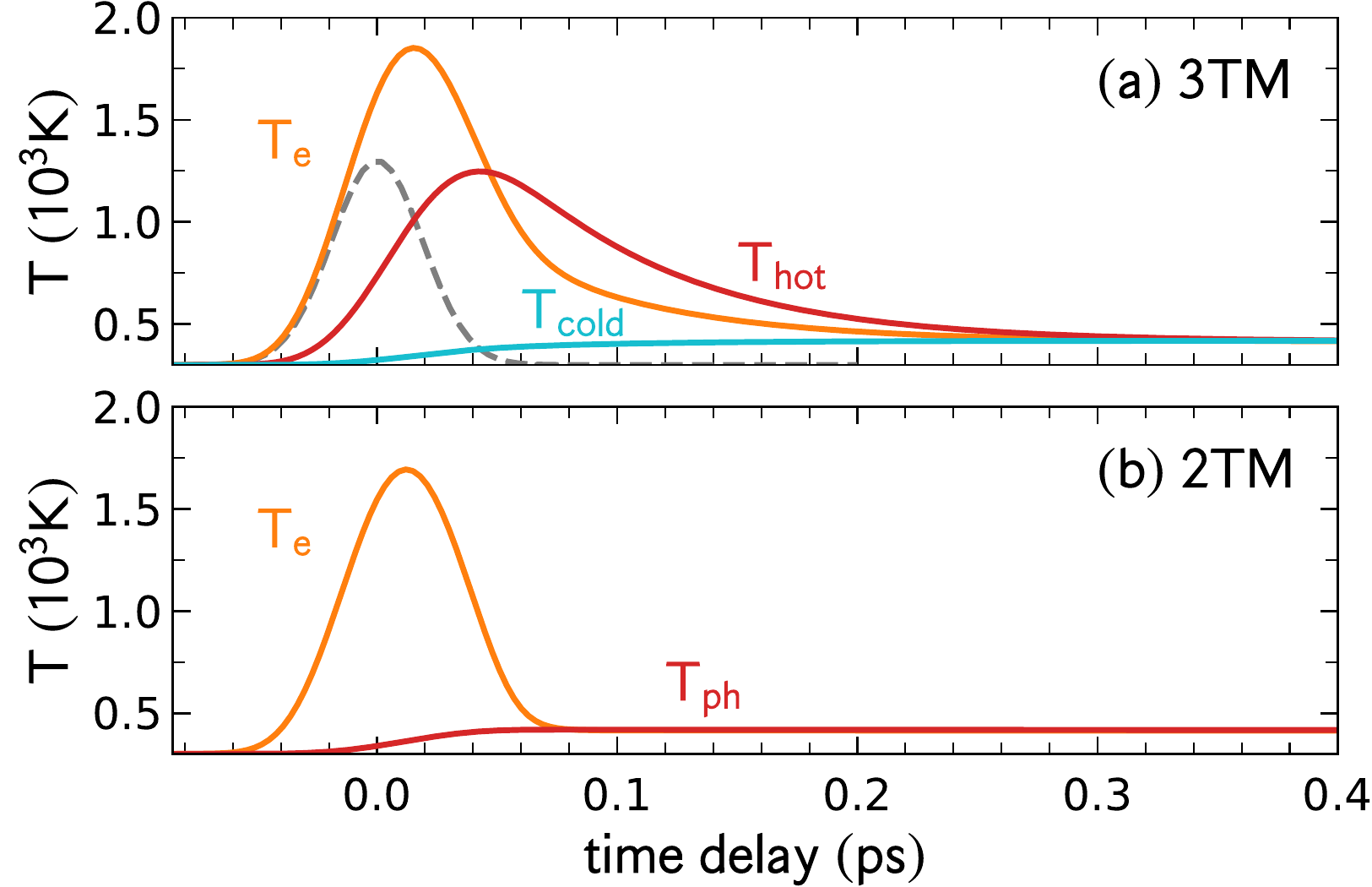}
\end{center}
\caption{(a) Time evolution of the characteristic effective 
temperatures $T_{\mathrm{e}}$, $T_{\rm hot}$, $T_{\rm cold}$
for the three-temperature model suitable for hot phonons.
The grey dashed line shows the pulse profile,
with the pulse duration of 45\, fs
and an absorbed fluence of 12\,J/m$^2$.
(b) Similar as in panel (a) within the assumption of
thermal distribution (two-temperature model) of the lattice modes as described
with Eqs.\,(\ref{eq:el2}) and (\ref{eq:eph2}).
Partial readaptation from Ref. \cite{ncdc}.
}
\label{f-time}
\end{figure}

The time evolution of the three characteristic temperatures
$T_{\rm e}$, $T_{\rm hot}$, $T_{\rm cold}$
(Fig.  \ref{f-time}a) shows a remarkable increase
of the effective temperature of the hot modes $T_{\rm hot}$,
which reaches a peak $T_{\rm hot}^{\rm max}\approx 1250$ K at $t^* \approx 50$\,fs
with a small delay compared to the time behavior
of the electronic temperature $T_{\rm e}$.
The remnant cold phonon modes follow, on the other hand,
a completely different behavior with a slow monotonic increase
towards the final equilibrium state at $t \ge 0.3-0.4$ ps
where all the degrees of freedom are thermalized with each other.

Such a behavior, showing a clear hot-phonon scenario, can be compared
with the isotropic case where the energy from the electronic degrees of freedom
is transferred in a equal way to {\em all} the lattice modes, without
a preferential channel, and one can thus define a unique phonon temperature
$T_{\rm ph}$ for all the modes\,\cite{bib:allen87}.
The coupled equations of the two temperature model can be obtained by setting $T_{\rm hot}=T_{\rm cold}\equiv T_{\rm ph}$ in Eqs.\,\eqref{eq:el}-\eqref{eq:ph}:
\begin{eqnarray}
C_{\rm e}
\frac{\partial T_{\rm e}}{\partial t}
&=&
S(z,t)
+
\nabla_z(\kappa\nabla_z T_{\rm e})
-
G_{\rm ph}(T_{\rm e}-T_{\rm ph})
,
\label{eq:el2}
\\
C_{\rm ph}\frac{\partial T_{\rm ph}}{\partial t}
&=&
G_{\rm ph}(T_{\rm e}-T_{\rm ph})
,
\label{eq:eph2}
,
\end{eqnarray}
where $C_{\rm ph}=C_{\rm hot}+C_{\rm cold}$
and $G_{\rm ph}=G_{\rm hot}+G_{\rm cold}$.
The time evolution of $T_{\rm e}$, $T_{\rm ph}$
is displayed in Fig. \ref{f-time}b. In the absence of the hot-phonon bottleneck, a much faster thermalization between the electrons and the lattice degrees of freedom is obtained,
within a time-scale of $t \sim 0.08$ ps, while the average phonon temperature $T_{\rm ph}$ reaches the maximum of about 420\,K.

\section{Time-resolved lattice dynamics}
\label{s-latticedynam}

As discussed above, the hot-phonon scenario is characterized by the regime
$T_{\rm hot} \gg T_{\rm cold}$, where the population of the hot-lattice modes
is singled out and is significantly larger than
the population of the other cold modes.
Equipped with the detailed input from the \emph{ab initio} calculations,
we discuss now how this scenario in MgB$_2$ can be revealed directly
in the dynamical lattice properties, i.e., in the amplitude
of the mean square lattice displacements, resolved for
atomic species and for different directions;
and more strikingly in the degree of {\em correlation} of the
atomic motion.
The peculiar properties of hot phonons in MgB$_2$
rely on the $E_{2g}$ symmetry of the strongly coupled
hot modes, in particular: ($i$) the pure in-plane B character of the
$E_{2g}$ modes; 
($ii$) the counter-phase motion of the two B atoms per cell
that move in opposite direction, as depicted in Fig. \ref{f-mgb2}a.

The evaluation of the mean square lattice displacements
based on the force-constant-model calculations has been detailed in Refs.\,\cite{jeong03,campi}.
Here, we extend such computation by using fully \emph{ab initio} methodology and, more importantly, by introducing the hot-phonon scenario
and specifying a different temperatures for each relevant modes.
More explicitly, we can define the {\rm projected} mean-square lattice displacement $\sigma^2(i_\alpha)$ as
\begin{equation}
\sigma^2(i_\alpha) = \langle [{\bf u}_i \cdot \hat{\bf r}_\alpha]^2 \rangle,
\end{equation}
where ${\bf u}_{i}$ is the lattice displacement of atom $i$ from
its average position and $\hat{\bf r}_\alpha$ is
the unit vector pointing along the direction $\alpha=x,y,z$.
On the microscopic ground we can write
\begin{equation}
\sigma^2(i_\alpha) =
\frac{\hbar}{N} \sum_{{\bf q},\mu}
\frac{|\epsilon_{{\bf q},\mu}^{i\alpha} |^2}
{M_i \omega_{{\bf q},\mu}}
\left[
\frac{1}{2}+b\left(
\frac{
\omega_{{\bf q},\mu}
}{
T_{{\bf q},\mu}
}
\right)
\right],
\label{e-s2}
\end{equation}
where $N$ is
the total number of ${\bf q}$-points considered in the phonon
Brillouin zone,
$M_i$ is the atomic mass of atom $i$,
$\omega_{{\bf q},\mu}$ is the frequency of a phonon mode
of branch $\mu$ with momentum ${\rm q}$,
$T_{{\bf q},\mu}=T_{\rm hot}, T_{\rm cold}$ is the appropriate
temperature for such mode,
and $\epsilon_{{\bf q},\mu}^{i\alpha}$ is the component of the corresponding
eigenvector $\hat{\epsilon}_{{\bf q},\mu}$ describing the
displacement of the $i$ atom along the $\alpha$ direction.

\begin{figure}[!t]
\begin{center}
\includegraphics[width=0.8\textwidth]{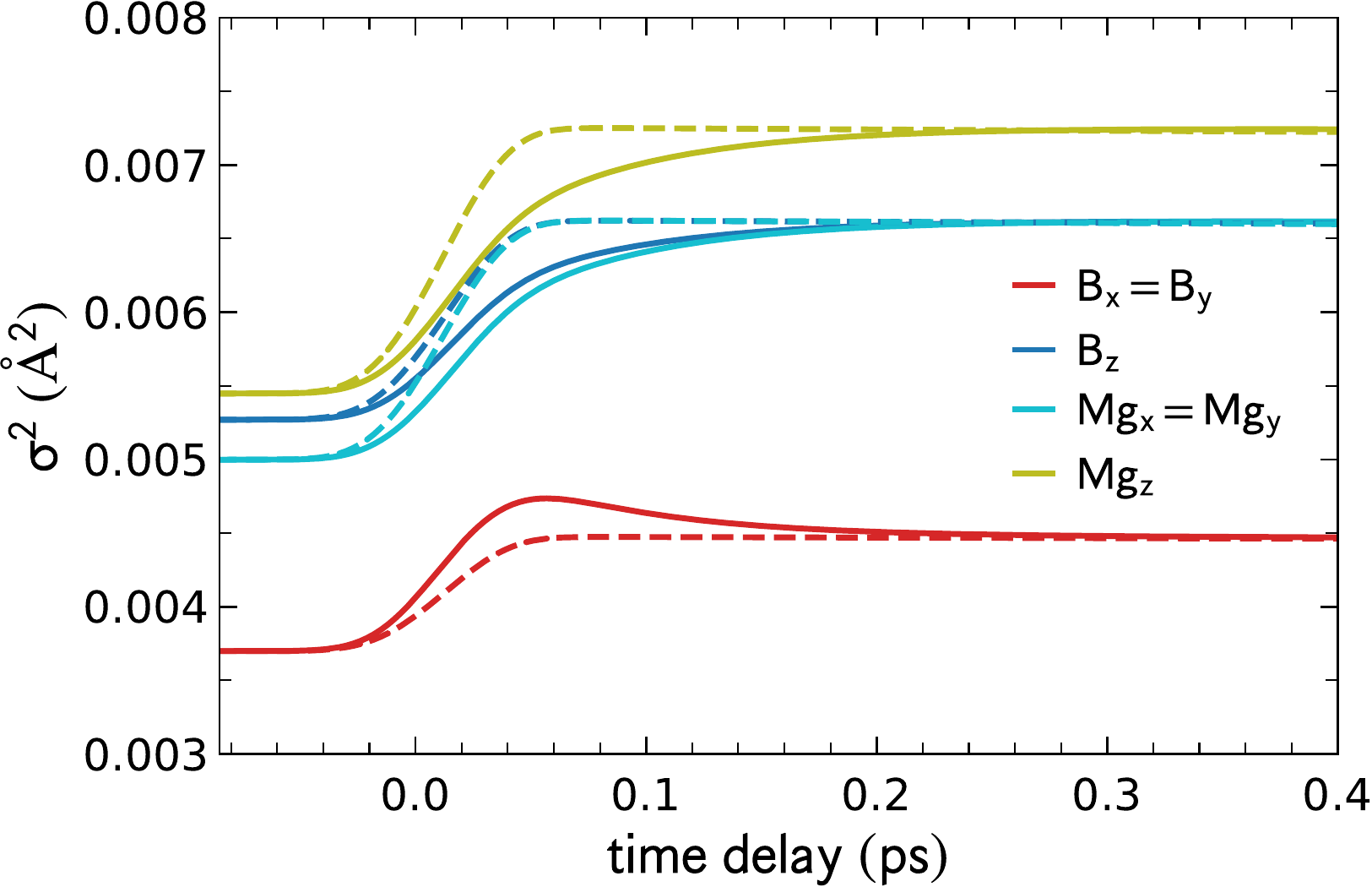}
\end{center}
\caption{Time evolution of the 
mean square lattice displacements $\sigma^2$
for each atom and for different axis directions
in the case of the hot-phonon scenario, described by Eqs.\,(\ref{eq:el})-(\ref{eq:ph}) (full lines). The corresponding results for the thermal phonon distribution [described by Eqs. (\ref{eq:el2})-(\ref{eq:eph2})] are shown with the dashed lines.
}
\label{f-s2t}
\end{figure}

The behavior of the mean square lattice displacements
resolved for each atom and for different axis directions in the case of hot non-thermal phonon distribution is shown
in Fig. \ref{f-s2t} (solid lines).
As expected,  compared with the case of a thermal phonon distribution described
by Eqs.\,(\ref{eq:el2})-(\ref{eq:eph2}) (dashed lines), we do not see any remarkable difference for the phonon modes with Mg character and with out-of-plane lattice displacements,
apart from a slower thermalization in the case of hot phonons,
due to the longer storage of energy in the hot-phonon modes.
A slight effect of the presence of hot phonons can be detected in the behavior
of the mean square in-plane B lattice displacements which shows
a non-monotonic behavior
with a peak value larger than the final equilibrium state,
at $t^*=50$\,fs, in fair accordance with the peak of  $T_{\rm hot}$.

A more striking signature
of the hot-phonon scenario can be detected
from the analysis of the correlated lattice dynamics.
More explicitly, we focus
on  the mean-square {\em relative} displacement of
atomic pairs projected onto the vector joining the atom
pairs\,\cite{jeong03,campi}:
\begin{equation}
\sigma^2_{ij} = \langle [({\bf u}_i-{\bf u}_j)\cdot
\hat{\bf r}_{ij}]^2 \rangle ,
\end{equation}
where ${\bf u}_i$ and ${\bf u}_j$ are the lattice displacements of
atoms $i$ and $j$ from their average positions, $\hat{\bf r}_{ij}$
is the unit vector connecting atoms $i$ and $j$.

This physical quantity can be evaluated microscopically as:
\begin{eqnarray}
\sigma^2_{ij} &=& \frac{\hbar}{N} \sum_{\bf q,\mu}
\left[\frac{1}{2}
+
b\left(
\frac{
\omega_{{\bf q},\mu}
}{
T_{{\bf q},\mu}
}
\right)
\right]
\Bigg\{
\frac{|\hat{\epsilon}^i_{{\bf q},\mu} \cdot \hat{\bf r}_{ij}|^2}
{M_i\omega_{\bf q,\mu}}
+\frac{|\hat{\epsilon}^j_{{\bf q},\mu} \cdot \hat{\bf r}_{ij}|^2}
{M_j\omega_{\bf q,\mu}}
\nonumber\\
&&
-\frac{2\mbox{Re}\left[
(\hat{\epsilon}^i_{{\bf q},\mu} \cdot \hat{\bf r}_{ij})
(\hat{\epsilon}^{j*}_{{\bf q},\mu} \cdot \hat{\bf r}_{ij})
\mbox{e}^{i{\bf q}\cdot {\bf r}_{ij}}
\right]}{\omega_{\bf q,\mu}\sqrt{M_iM_j}}
\Bigg\}.
\label{e-corr}
\end{eqnarray}

It is useful to quantify the degree of correlation
by introducing a convenient dimensionless correlation factor
$\rho_{ij}$ defined as
\begin{eqnarray}
\sigma^2_{ij}
&=&
 \sigma^2(i_j) + \sigma^2(j_i)
-2\sigma(i_j)\sigma(j_i)\rho_{ij},
\label{e-rho}
\end{eqnarray}
where 
\begin{eqnarray}
\sigma^2(i_j)
&=&
\langle [({\bf u}_i\cdot\hat{{\bf r}}_{ij}]^2 \rangle
\nonumber\\
&=&
 \frac{\hbar}{N} \sum_{\bf q,\mu}
\left[\frac{1}{2}
+
b\left(
\frac{
\omega_{{\bf q},\mu}
}{
T_{{\bf q},\mu}
}
\right)
\right]
\frac{|\hat{\epsilon}^i_{{\bf q},\mu} \cdot \hat{\bf r}_{ij}|^2}
{M_i\omega_{\bf q,\mu}}
.
\end{eqnarray}
The inter-atomic correlation factor
can be thus calculated as\cite{campi}
\begin{eqnarray}
\rho_{ij}
&=&
\frac{\sigma^2(i_j) + \sigma^2(j_i)
-\sigma^2_{ij}}{2\sigma(i_j)\sigma(j_i)}.
\label{e-rho2}
\end{eqnarray}
Positive values of correlation factor $\rho_{ij} > 0$ describe a situation where
the couple of atoms $i$, $j$ move in phase, so that the resulting value of
$\sigma^2_{ij}$ is smaller than for the uncorrelated case. On the
other hand, a predominance of counter-phase atomic vibrations
is expected to result in a negative correlation factor $\rho_{ij} < 0$.

Equations (\ref{e-s2}) and (\ref{e-rho2}) provide fundamental tools
for investigating the effects of the hot-phonon scenario
on the time evolution of the correlated lattice dynamics.
At the initial thermal equilibrium at room temperature we find positive values
$\rho_{\rm B-B}\approx 0.24$ \cite{noteexpcorr},
$\rho_{\rm B-Mg}\approx 0.22$,
reflecting a slight dominance of the acoustic (in-phase) modes
in contributing to the lattice dynamics.
\begin{figure}[!tb]
\begin{center}
\includegraphics[width=0.8\textwidth]{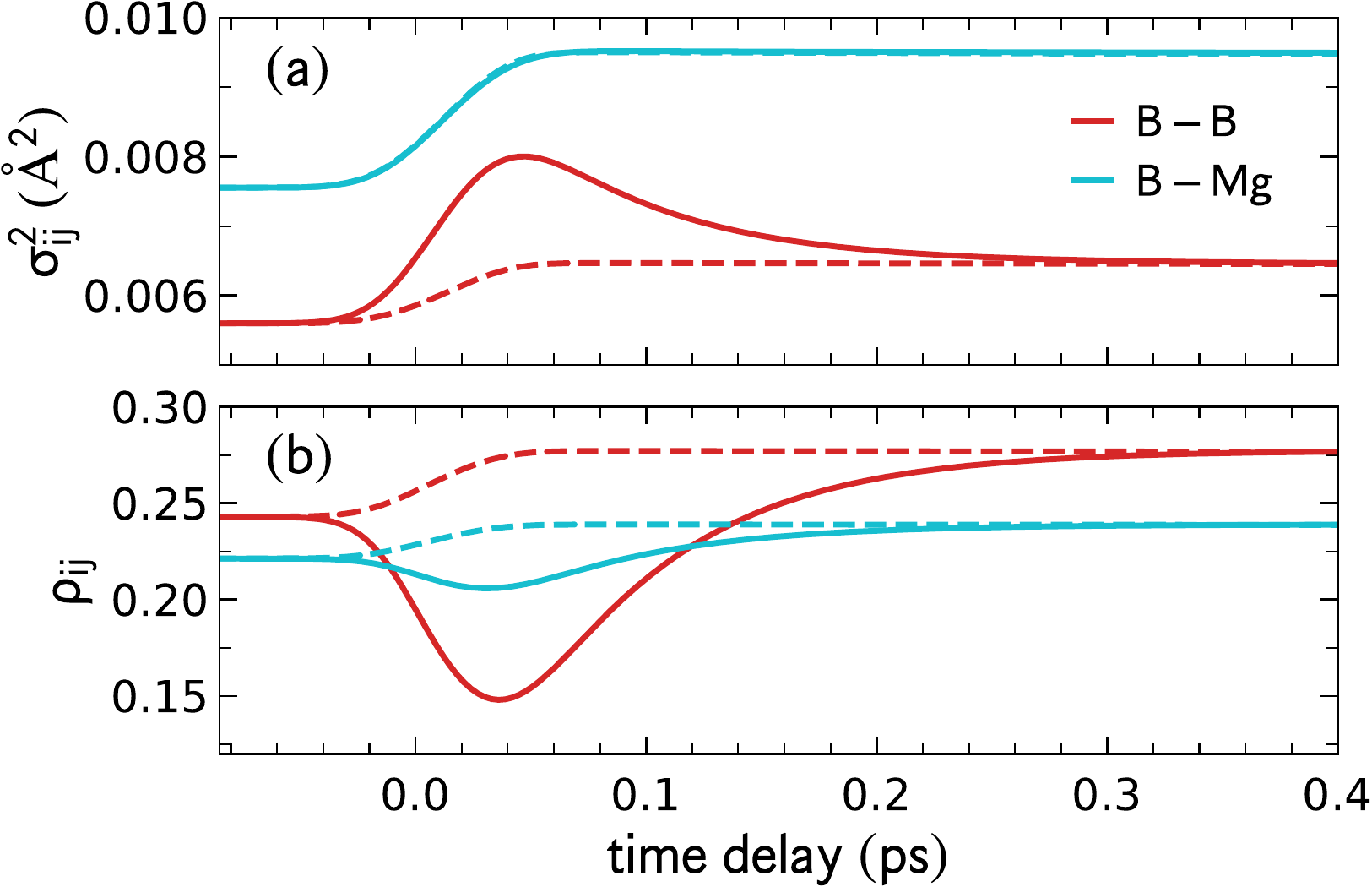}
\end{center}
\caption{(a) Time evolution of the 
mean-square relative lattice displacement $\sigma_{ij}^2$ as defined by Eq. (\ref{e-corr}),
for the in-plane nearest neighbor B-B bond, and for the nearest neighbor B-Mg bond.
Solid lines show the hot-phonon scenario, whereas dashed lines represent the
time behavior assuming a a thermal phonon distribution as modelled
by Eqs. (\ref{eq:el2})-(\ref{eq:eph2}).
(b) Corresponding correlation functions $\rho_{ij}$.
}
\label{f-corr}
\end{figure}
The time dependence of
the mean-square {\em relative} displacement for
the in-plane boron-boron pair, $\sigma_{\rm B-B}$,
and for the interplane boron-magnesium one
$\sigma_{\rm B-Mg}$ is reported in Fig. \ref{f-corr}a.
The inter-atomic boron-magnesium lattice displacement (blue full line)
does not show any significant feature,
just a monotonic increase between the two equilibrium states,
as in the absence of hot phonons (blue dashed line).

A definitive fingerprint of the crucial role
of the hot-phonon is provided by
the time evolution of the mean-square relative
in-plane boron-boron displacements $\sigma_{\rm B-B}$ as shown in Fig. \ref{f-corr}a.
For the assumption of a thermal phonon distribution (red dashed line), $\sigma_{\rm B-B}^2$ increases monotonically
obeying to a thermal heating of the lattice (similar as $\sigma_{\rm B-Mg}^2$).
Within the hot-phonon scenario, however,
the in-plane nearest neighbor mean-square relative lattice displacement  $\sigma_{\rm B-B}^2$
presents a remarkable peak at the highest $T_{\rm hot}$ achievable at $t^*=50$\,fs,
with a following decrease towards a new equilibrium conditions (red full line).
Such a peak is a direct consequence of a dominance
of a larger population of the in-plane boron modes
with the $E_{2g}$ symmetry.

The relevance of hot-phonon physics is revealed in a even more
striking way in the analysis of the lattice displacement
correlation factor $\rho_{ij}$.
As shown in Fig. \ref{f-corr}b, the onset of $E_{2g}$ hot phonons in MgB$_2$
is reflected in a direct way in a remarkable anomaly
in the time evolution of the relative boron-boron in-plane lattice
correlation, with a sharp decrease
of $\rho_{\rm B-B}$ in the first 50\,fs.
The presence of hot-phonon modes with
pure in-plane boron character is visible also
in the interplane boron-magnesium correlation factor
$\rho_{\rm B-Mg}$, although in a weaker way.
Such anomalies are peculiar of the hot-phonon scenario
and are not predicted to show when the lattice modes
are assumed to obey a thermal distribution (dashed lines).

Our analysis provides thus a simple and straightforward way to probe
the presence of hot phonons in MgB$_2$ directly in the
analysis of the lattice dynamics properties,
as it can be revealed in time-resolved pump-probe experiments.
More precisely, we predict a marked anomaly in the
time behavior of lattice correlation factor $\rho_{\rm B-B}$
occuring at the delay time $t^*$ when the hot-phonon lattice modes
reach their largest population.
The time scale over which such anomaly disappears prompts
also a possible direct procedure for evaluating experimentally
the time scale when {\em all} the lattice degrees of freedom reach
their final thermalization.

\section{Wider perspective on other materials and experimental observations}
\label{s-wider}

The approach described in the present work suggests a possible way
to detect hot phonons not in an indirect way through the effects
on the electronic properties but directly in the lattice dynamics,
exploiting the consequences in the correlation
of inter-atomic motion.
In the specific case of MgB$_2$, the candidate modes for hot phonons
are the ${\bf q}=0$ optical $E_{2g}$ ones that correspond to
out-of-phase in-plane motion of the two B atoms.
The large non-thermal population of these modes gives rise thus
to a marked anomaly in the in-plane correlation factor $\rho_{\rm B-B}$
with respect to the steady conditions, as depicted in Fig. \ref{f-corr}.

As we stressed above, such approach is not specific of MgB$_2$ but it is quite general
and it can be usefully applied to other materials.
The overall scheme, for a given
material candidate to hot-phonon physics, can be summarized as follows
($i$) identifying the lattice modes with particularly large
el-ph coupling, likely thus to sustain hot-phonon physics;
($ii$) analyzing how the eigenvectors of such modes
affect the inter-atomic correlated motion;
($iii$) thus investigating and predicting specific anomalies
in the appropriate correlation factors, validating such prediction
with the comparison with a overall thermal heating of the lattice
with no hot phonons included.
The most straightforward generalization is probably graphene,
where hot phonons have been assessed in many ways, and
which shares many similarities with MgB$_2$.
The $E_{2g}$ modes of the hexagonal B plane MgB$_2$ are the same 
$E_{2g}$ modes of the hexagonal C plane that
get hot in graphene.
We can hence predict that a similar anomaly can be observable
also for graphene.

\begin{figure}[!tb]
\begin{center}
\includegraphics[width=0.8\textwidth]{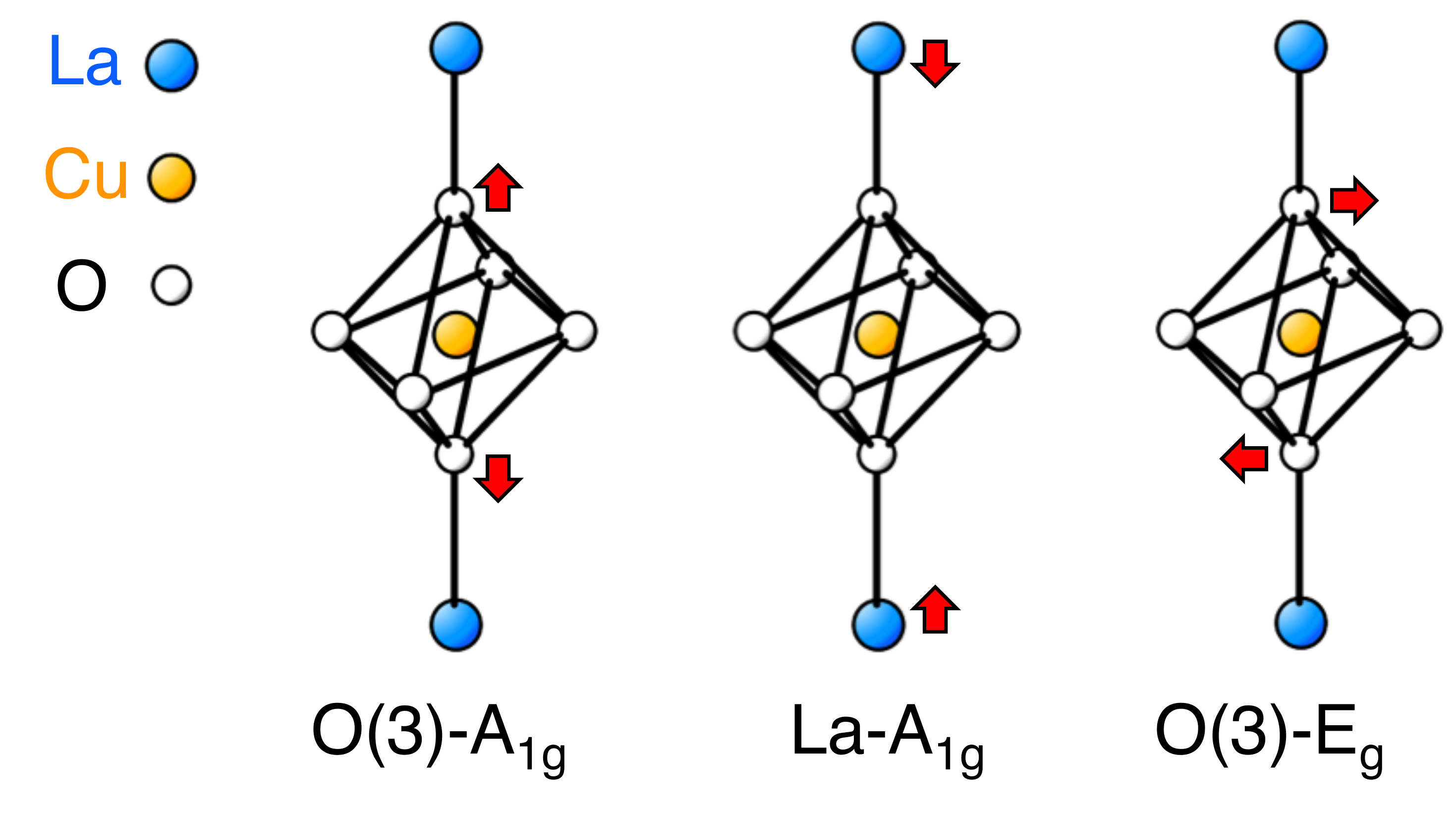}
\end{center}
\caption{Main optical lattice modes of La$_{2-x}$Sr$_x$CuO$_4$ displaying sizable
el-ph coupling and which are thought to support hot-phonon physics.
Sr is a random substitutional atom for La and it is not shown here.
La$_{2-x}$Sr$_x$CuO$_4$ has an layered perovskite structure with apical-oxygen
distance larger than the in-plane oxygens.
}
\label{f-cuprates}
\end{figure}

A similar approach is expected to work also for more complex materials
with several atoms in unit cell.
A suitable class of compounds, along this perspective,
is the family of superconducting cuprates, as for instance
the representative system La$_{2-x}$Sr$_x$CuO$_4$ 
where the occurence of hot-phonon physics has been already discussed in literature \cite{johnson17}.
The lattice modes that get hot in this compound are predicted to belong mainly to
the $A_{1g}$ symmetry, involving out-of-plane displacements either of the La atoms
or of the apical oxygen atoms [often labelled as O(3)] \cite{mansart13}. A residual el-ph coupling
also involves $E_g$ modes of the O(3) oxygens.
A sketch of the relevant modes is provided in Fig. \ref{f-cuprates}.
A time-resolved assessement of the correlated inter-atomic motion can be a poweful tool
also in this case since it could distinguish the role of the different modes
in the La-La or in the O($i$)-O($i$) inter-atomic correlations ($i=1,2,3$).
Note that the different contribution of the in-plane or apical oxygens
in the inter-atomic lattice correlations, as probed for instance by
the inter-atomic probability distribution function, could be easily reveled
due to the different bond lengths of the different oxygens \cite{cohen89,pickett89}.

Other more complex materials with several atoms for cell, sustaining hot-phonon physics,
could be in principle devised.
In such a framework,
on the theoretical side, one cannot exclude that in-phase motions can be favoured,
resulting in a {\em positive} anomaly in a specific inter-atomic correlation function,
when the out-of-phase motion
of a subset of atoms is accompanied by an in-phase motion
of another subset of atoms.

From the experimental point of view, it should be noticed that the most accurate
probes of the inter-atomic correlated motion under equilibrium conditions
are presently provided by neutron diffraction measurements
that are probably not best suited for time-resolved femtosecond resolution.
Our theoretical proposal does not however rely on a specific
experimental setup.
It should be noticed indeed that, even
in the steady neutron diffraction measurements, raw data are collected
in the momentum space, and later translated in the real-space
by Fourier transform.
In this perspective ultrafast electron diffuse scattering \cite{trigo10,chase16,stern18},
as well as time-resolved resonant inelastic X-ray scattering \cite{ament11,mitrano20},
represent two promising alternative techniques
which are rapidly ramping up and
constantly expanding their accuracy and limits.

\section{Conclusions}
\label{s-conclusions}

In this paper, using first-principle calculations,
we have analyzed the presence of the hot-phonon physics in MgB$_2$ and its effect on the time evolution of lattice dynamics
in a typical pump-probe experiment.
We have shown that hot phonons in MgB$_2$,
with their characteristic $E_{2g}$ symmetry that implies an in-plane counter-phase
motion of the boron atoms,
can be directly traced down in the analysis of the atom-resolved
mean-square lattice displacements, and more evidently
in the analysis of the mean square relative
lattice displacements $\sigma_{ij}^2$ and lattice correlation factors $\rho_{ij}$.
Even though we apply our investigation to the specific case of MgB$_2$, the analysis here described
is quite generic and it can be extended to a wide
range of materials (semiconductors, metals or semimetals)
sustaining hot phonons, since they are commonly established at high-symmetry points
of the Brillouin zone, with well-defined and well-known
symmetry and atomic contents.
Promising compounds might be for instance
single-layer and multi-layer graphene\,\cite{novko2019,caruso20}, transition metal dichalcogenides\,\cite{caruso21},
black phosphorous\,\cite{seiler21,zhan20}, hole-doped diamond\,\cite{boeri04,giustino07}
and underdoped cuprates \cite{mansart13,johnson17}.
Our work provides thus a guideline for future direct experimental verification
and characterization of hot-phonon mechanisms
in several families of compounds.

\section*{Acknowledgements}
D.N. acknowledges financial support from the Croatian Science Foundation (Grant no. UIP-2019-04-6869) and from the European Regional Development Fund for the ``Center of Excellence for Advanced Materials and Sensing Devices'' (Grant No. KK.01.1.1.01.0001).




\bibliography{biblio2.bib}

\nolinenumbers

\end{document}